\documentclass[a4,12pt]{revtex4-2}
\usepackage[utf8]{inputenc} \usepackage[T1]{fontenc}
\usepackage{times}

\usepackage{amsmath}
\usepackage{bm}
\usepackage{amssymb,latexsym,mathrsfs}

\usepackage{url}
\usepackage{color}

\usepackage{tikz}    
\usepackage{xcolor}  

\usepackage[sf,pagestyles]{titlesec}
\titleformat{\section} {\normalfont\sffamily\bfseries} {\thesection}{1em}{}

\long\def\comment#1{}

\usepackage[bookmarks=false,breaklinks]{hyperref}

\definecolor{royalbluecs}{rgb}{0.2, 0.2, 1}

\hypersetup{linktoc=all, 
    colorlinks=true, linkcolor={royalbluecs},
    citecolor={royalbluecs}, urlcolor={royalbluecs}}

\definecolor{lime}{HTML}{A6CE39}
\DeclareRobustCommand{\orcidicon}{%
	\begin{tikzpicture}
	\draw[lime, fill=lime] (0,0) 
	circle [radius=0.16] 
	node[white] {{\fontfamily{qag}\selectfont \tiny ID}};
	\draw[white, fill=white] (-0.0625,0.095) 
	circle [radius=0.007];
	\end{tikzpicture}
	\hspace{-2mm}
}

\foreach \x in {A, ..., Z}{%
	\expandafter\xdef\csname orcid\x\endcsname{\noexpand\href{https://orcid.org/\csname orcidauthor\x\endcsname}{\noexpand\orcidicon}}
}

\usepackage{etoolbox}
\makeatletter
\patchcmd{\frontmatter@RRAP@format}{(}{}{}{}
\patchcmd{\frontmatter@RRAP@format}{)}{}{}{}
\renewcommand\Dated@name{} 
\makeatother

\normalsize\normalfont

\begin{document}

\title{\textsf{From maximum force to the field equations of general relativity\\ -- and implications}}
%

\author{ \ \\
Arun Kenath, {CHRIST (deemed to be University), Bangalore, India, \email{kenath.arun@christuniversity.in}, ORCID 0000-0002-2183-9425 \orcidA{}} \\
\ \\
Christoph Schiller, {Motion Mountain Research, Munich, Germany, \email{cs@motionmountain.net} (corresponding author), ORCID 0000-0002-8188-6282 \orcidB{}} \\
\ \\
C. Sivaram, {Indian Institute of Astrophysics, Bangalore, India, \email{sivaram@iiap.res.in}, ORCID 0000-0001-7423-4078 \orcidC{}}  \\
\ \\
}

\begin{abstract} 
\noindent \ \hfil \par
\noindent {\bf\textsf{Abstract}}\par\noindent
There are at least two ways to deduce Einstein's field equations from the principle of maximum force $c^4/4G$ or from the equivalent principle of maximum power $c^5/4G$. Tests in gravitational wave astronomy, cosmology, and numerical gravitation confirm the two principles. Apparent paradoxes about the limits can all be resolved. Several related bounds arise. The limits illuminate the beauty, consistency and simplicity of general relativity from an unusual perspective.\hfil
\end{abstract} 

\bigskip

\bigskip

\emph{\hspace*{0em}Essay written for the Gravity Research Foundation 2022 Awards for Essays on Gravitation. \hfil}

\emph{\hspace*{0em}Selected for Honorable Mention. \hfil}

\bigskip

\bigskip

\date{22 March 2022} 

\bigskip

\maketitle


\newpage
\section{History and experiments}

\noindent It is known since around 1887 that the invariant maximum speed $v_{\rm max}= c \approx 3.0 \cdot 10^{8} \rm \,m/s$ is valid for every physical system. Since 1905 it is known that maximum speed \emph{implies} special relativity.
Similarly, it is known since around 1973 that the invariant maximum force  
\begin{equation}
     F_{\rm max}=
	 {c^4}/{4G} \approx 3.0 \cdot 10^{43} \rm \,N \;\;,
	 \label{eq1}
\end{equation}
is valid for every physical system.
Power, the zeroth component of the force four-vector, has an equivalent limit 
\begin{equation}
     P_{\rm max}= 
	 {c^5}/{4G} \approx 9.1 \cdot 10^{51} \rm \,W \;\;.
	 \label{eq1b}
\end{equation}
Since 2002 it is known that maximum force or maximum power \emph{imply} the field equations of general relativity. 
Either limit allows exploring general relativity in an intuitive and productive way. 
The limits help to get an overview about the effects and the beauty of gravity, including curvature, horizons, black holes and gravitational waves.

It was Elizabeth Rauscher who mentioned maximum force for the first time, in 1973 \cite{rauscher}. 
The concept had been overlooked by Einstein and many researchers after him.  
She was followed by Treder \cite{treder}, Heaston
\cite{10.2307/24530850}, de Sabbata and Sivaram \cite{sab} and others
\cite{massa,Kostro:1999ue}.  After the turn of the millennium, the topic was explored in more detail, and the
factor $1/4$ was deduced by Gary Gibbons
\cite{gibbons} and confirmed by Schiller \cite{s1,csmax}. Recently, interest in maximum force has increased
\cite{hogan,barrow1,ong2,abo,Ong:2018xna,barrow2,barrow3,b1,MYNEWPRD,Ong:2018xna,Gurzadyan:2021hgh,michael2,siva,Dadhich:2022yuk}.

Several arguments lead to a maximum force value. First of all, the limit follows from $F=ma$. The acceleration of (the front of) a body of length $l$ is limited by $a\leq c^2/l$ \cite{Taylor1983LimitationOP}. As a consequence, the force on a body of mass $m$ is limited by $F\leq c^2 (m/l)$. The largest mass per length ratio $m/l$ arises for a Schwarzschild black hole, with a value $c^2/4G$. This yields a maximum force value $F_{\rm max}={c^4}/{4G}$.

Equivalently, force is an energy per length: a force acting along a path -- such as a force on a spring -- deposes an energy along the path. 
The highest possible energy per length ratio occurs when a Schwarzschild black hole of energy $Mc^2$ is deposed over its own diameter $4GM/c^2$.  
This ratio again yields a maximum force value $c^4/4G$.

Also the force 
produced on a test mass $m$ by a Schwarzschild black hole is limited.
When a test mass is lowered, using a string, towards the horizon, the value of the force of gravity can be calculated \cite{gibbons,ohanian,haye,MYNEWPRD}. 
At the horizon, the expression increases beyond all limits.
However, the test mass has a certain size, which is surely larger than its own Schwarzschild radius.
This minimum size of the test mass yields a smallest possible value for its distance to the horizon. 
This smallest distance prevents the force of gravity from ever exceeding the maximum value.
In fact, the same force limit arises when considering the force of gravity between any two black holes. 
Also their attraction never exceeds the maximum value \cite{gibbons,MYNEWPRD}.

Physically, the maximum force is equivalent to the maximum power or luminosity
    $ P_{\rm max}= c \, F_{\rm max} \approx 9.1 \cdot 10^{51} \rm \,W$.
This huge value corresponds to about $50\,700$ solar masses per second. 
Probably the first to investigate maximum power was Sciama, again in 1973 \cite{sciama1}.
A hint in the Gravitation textbook, Sivaram \cite{mtw,lastsivaram} and other authors followed
\cite{massa,hogan,Kostro:2000gw,Cardoso:2018nkg,Ong:2018xna,abo,Gurzadyan:2021hgh}. 
The factor $1/4$ arose later, in parallel with maximum force.

Both the maximum force and the maximum power values are so large that they do not arise in everyday life or in experiments. 
Indeed, both limit values are only relevant near black hole horizons.  
And even though no force value close to the maximum force has ever been measured, the situation differs for maximum power.
The most powerful energy sources in nature are black hole mergers.
Among these, the most powerful event observed by the LIGO and Virgo gravitational wave detectors reached an instantaneous emitted 
    power -- during about a millisecond -- of $230\pm80$ solar masses per second \cite{ligo1}, or $0.46\pm0.16\,\%$ of the maximum power value. 
The well-known 2019 black hole merger emitted up to $207\pm50$ solar masses per second \cite{ligo2}. 
Therefore, observations with gravitational waves are $2$ orders of magnitude away from possible falsification. 
Future, space-based detectors will do better.
But the chances for falsification are low:
    even the luminosity of the full universe does not and never did exceed the value $c^5/4G$.

\section{A first derivation of the field equations from maximum force}

\noindent There is a common reticence to use the concept of force in general relativity.
Nevertheless, force, with its usual definition as change of momentum, $F=dp/dt$, can be used without restrictions. 
If preferred, the more intuitive concept of maximum power can be used instead.

The invariance of the speed of light implies that around a mass, space is \textit{curved}. 
Spatial curvature is confirmed by observations during solar eclipses. 
Curvature is also predicted by maximum force: the limit is only achieved at black hole horizons, and such horizons only arise around masses, and only due to the curvature of space. %
In short, maximum force implies that masses bend vacuum: vacuum is elastic.

The elasticity of a material is described by the \textit{shear modulus}. 
The shear modulus also determines the \textit{shear strength}, i.e., the maximum shear that a material can endure before plastic deformation and ripping apart. 
The two quantities are related by a factor of order $1$.
We now imagine vacuum to be a material.
The elastic constant $c^4/8 \pi G$ of the vacuum determines, within a factor $2 \pi$, the maximum force $c^4/4G$ that the vacuum can endure before plastic deformation. (We will explore below what happens in this extreme case.)

General relativity  describes the \textit{elastic} behaviour of space. 
The maximum force value and the elastic constant of vacuum imply a linear relation between spatial curvature and energy density. 
When this relation is generalized to arbitrary observers and thus to general coordinate systems, the field equations arise directly \cite{siva,MYNEWPRD}.
To see this, one starts with force $F_{\rm max}$ acting over an area $A$. 
This yields a stress, or energy density, given by
\begin{equation}
	\epsilon=\frac{c^4}{4G}   \, \frac{1}{A}         \;\;.           
	 \label{eq2}
\end{equation}
In the case of a spherical surface of radius $a$, the area $A=4 \pi a^2$ implies a uniform curvature $R=1/a^2$. 
Therefore, energy density and curvature are proportional, as expected from elasticity: $\epsilon = ({c^4}/{16 \pi G}) \, R$. 
In the case of a general coordinate system, the energy density $\epsilon$ is described by the energy–momentum tensor $T_{\mu\nu}$ (with components of pressure, density, etc.) and the contracted curvature $R/2$ becomes the Ricci tensor $R_{\mu\nu} $.
This yields the intermediate result:
\begin{equation}
	T_{\mu\nu}    =\frac{c^4}{8 \pi G} \, R_{\mu\nu}    \;\;.    
	 \label{eq4}
\end{equation}
The conservation of $T_{\mu\nu} $, i.e., the vanishing of its covariant derivative, dictates that $R_{\mu\nu} $ be replaced by 
     the Einstein tensor  $G_{\mu\nu} = R_{\mu\nu}  -  g_{\mu\nu}  R/2  $,  
whose covariant derivative also vanishes, leading to 
\begin{equation}
	T_{\mu\nu}    =\frac{c^4}{8 \pi G} \, G_{\mu\nu}    \;\;.    
	 \label{eq4b}
\end{equation}
\noindent
These are Einstein's field equations of general relativity.
The derivation can be extended to include the cosmological constant.
In short, using an argument inspired by vacuum elasticity, maximum force implies the field equations.

\section{A second derivation of the field equations}

\noindent 
A second derivation of the field equations focuses on the properties of black hole horizons \cite{csmax,MYNEWPRD}. 
A black hole can be seen as matter in permanent free fall.
In other terms, every black hole horizon shows energy flow. 
Because a black hole horizon has a finite radius, at every point of the horizon, the maximum force value limits the energy flow through the horizon. 
%
The limit is called
the \emph{first law of black hole mechanics} \cite{Bardeen:1973gs,wald}.
The first law follows from and is equivalent to the fact that event horizons are surfaces showing maximum force at every point. 
(The first law also arises if one starts with maximum power instead of maximum force.)

The first law describes how, for a given horizon surface gravity, a change in horizon area induces a change in horizon energy: the law thus specifies the dynamics of horizons and shows that this dynamics is determined by the maximum force. %
This is analogous to special relativity, where the dynamics of massless radiation is determined by maximum speed.

In 1995, Jacobson showed \cite{jac} that the first law of horizon mechanics is equivalent to Einstein's field equations of general relativity, including the cosmological constant.
The result was confirmed by other authors \cite{paddy1,paddy2,isohor,hayward,oh}. %
The argument is twofold: 
the first law implies that horizons follow the field equations.
In addition, using a suitable coordinate transformation, 
     a horizon can be positioned at any desired location in space-time. %
This possibility implies that the dynamics of space-time, also far from any horizon,
     is equivalent to the dynamics of horizons.
As a result, the field equations are valid everywhere.

In short, maximum force or maximum power, together with the maximum speed, \textit{imply} the first law of horizon mechanics;
the first law in turn \textit{implies} the field equations. 

\section{The \textit{principle} of maximum force or power}

\noindent Every step in the previous two derivations of the field equations can be reversed. 
As a consequence, the field equations and maximum force, or maximum power, are {equivalent}. 
It is therefore correct to speak about the \textit{principle} of maximum force or about the principle of maximum power in general relativity. 
This is akin to speaking about the \textit{principle} of maximum speed in special relativity.

Special relativity follows from maximum speed -- in flat space.
The simplicity of the statement suggests that there are no modifications to special relativity in nature -- for flat space. 
Likewise, general relativity follows from maximum force -- in curved space.
The simplicity of the statement suggests that there are no modifications to general relativity in nature -- for curved space.

Indeed, the principle of maximum force implies that every test of general relativity \textit{near a horizon} is, at the same time, a test of maximum force. 
At present, the search for deviations from general relativity, especially near horizons, is a vibrant research field. 
But even the most recent observations of black hole mergers \cite{gtc3} and of the double radio pulsar PSR J0737–3039A/B \cite{doublepulsar} failed to find any deviation. %

The force and power limits also allow to test alternative theories of gravitation.
So far, theories that contradict experiments also appear to contradict the two limits, and vice versa \cite{abo,Wang:2020bjk,Vijaykumar:2020nzc,Le:2021hej,atazadeh}.

\section{Counter-arguments and paradoxes}

\noindent The statement of a maximum force $c^4/4G$ elicits attempts to exceed it.  
For example, one can ask whether Lorentz boosts allow exceeding the maximum force. 
But since a long time, textbooks implicitly show that this is impossible: both the acceleration value and the force value in the \emph{proper} frame of reference are not exceeded in any other frame \cite{pauli,moeller, Rebhan:1999}. %
For example, in the simple one-dimensional case, the boosted {acceleration} value is the proper acceleration value divided by $\gamma^3$, whereas the boosted force value is the same as the proper force value.

What happens if one adds two forces whose sum is larger than the maximum? 
If the forces act at different points, their sum is \textit{not} limited by the principle of maximum force.  
Any force is a momentum flow; the principle does not limit the sum of flows at different locations.
In the same way that adding speeds at different points in space can give results that exceed the speed of light, also adding forces at different points in space can give values exceeding the force limit. %
If, instead, the forces in question all act at a single point, the principle states that their sum \textit{cannot} exceed the maximum value \cite{MYNEWPRD}. 
The speed and force limits are \textit{local.}

A recent proposal for exceeding the maximum force \cite{jowsey} disregarded locality, as explained in reference \cite{newcs}. 
In fact, whenever one tries to exceed maximum force at a specific location, a horizon appears that prevents doing so,
because the principle of maximum force also implies the hoop conjecture \cite{thorne,Hod:2020yub}.

Why does the weakest interaction determine the highest force? 
Gravity's ``weakness'' is due to the smallness of typical elementary particle masses, and not to an intrinsic property of gravity \cite{Wilczek:2001ty}. 
In fact, all interactions lead to curvature. 
The relation between curvature and energy density described by $c^4/4G$ is independent of the type of interaction.

Another potential counter-argument arises from the issue of renormalization of $G$ in quantum field theory. 
Certain approaches to this issue \cite{Frolov:1996aj,Visser:2002ew,Volovik:2003kt,Hamber:2006sv} suggest that $G$ might change with increasing energy, and in particular that $G$ might increase when approaching the Planck energy. 
However, reasons for a fundamental impossibility that $G$ is renormalized were given by Anber and Donoghue \cite{Anber:2011ut,Donoghue:2019clr}.
So far, no hint for a change of $G$ with energy has been found. 
If future experiments ever find such such a change, maximum force is falsified. 

Several papers have claimed that the numerical factor in maximum force or power is $1/2$ instead of $1/4$ \cite{hogan,Ong:2018xna,abo,Gurzadyan:2021hgh}. 
However, the missing factor $1/2$ reappears either when distinguishing radius and diameter, or when the factor $2$ in the expression $E=2TS$, valid for black hole thermodynamics, is taken into account. %

Also the maximum power $c^5/4G$ leads to paradoxes. 
At first sight, it seems that the maximum power can be exceeded by combining two (or more) separate power sources that add up to a higher power value. 
However, at small distance from the two sources, their power values cannot be added, as this violates locality. 
And seen from a large distance, where such an addition might approximately be possible,  the two sources will partially absorb each other's emission and thus avoid exceeding the power limit. %
 
Another recent theoretical attempt \cite{jow2} to invalidate the power limit in star explosions made use of an expansion front speed larger than $c$. 
But front speeds are energy speeds; they cannot exceed $c$. 
Maximum power remains valid also in explosions.

In short, no confirmed counter-example to maximum speed, maximum force or maximum power has yet been found. 
Obviously, as usual in physics, falsification is still possible in the future -- though rather improbable.

\comment{
\section{From maximum force to inverse square gravity}

\noindent 
Maximum force allows an extremely simple derivation of inverse square gravity, using a few lines of simple algebra \cite{MYNEWPRD}.
First of all, maximum force $c^4/4G$ is a limit on energy per length.
This implies that there is a maximum possible energy value that can be enclosed in a sphere.  %
The existence of such a limit is called the hoop conjecture \cite{thorne,Hod:2020yub}.
The hoop conjecture thus follows from maximum force.

Combining the hoop conjecture with the size-acceleration limit of special relativity yields the first law of horizon mechanics; the first law was already mentioned above \cite{Bardeen:1973gs,wald} as a consequence of maximum force.

Inserting the relation $E=Mc^2$ and the surface expression of a sphere -- valid for flat space and thus away from any horizon -- into the first law immediately yields
\begin{equation}
	 a= \frac{MG}{r^2} \;\;.
\end{equation}
Therefore, in flat space, Newton's inverse square law of gravity is a direct consequence of maximum force and maximum speed. 
The limits indeed simplify the exploration of general relativity.
}

\section{Related limits in nature}

\noindent 
Similar to maximum force and maximum power, also the limit for mass flow rate given by $(dm/dt)_{\rm max}=c^3/4G \approx 1.0 \cdot 10^{35} \, \rm kg/s $ holds at every point in space-time. %
Again, the limit is realized only at horizons. 
However, even the most recent numerical simulations by Cao, Li and  Wu find no evidence that contradicts the limit \cite{Cao:2021mwx}. 

Any source emitting gravitational waves with maximum power yields,
at distance $r$, a limit on the curvature of space. As mentioned above, the limit has
not been exceeded in any observation by LIGO or Virgo.

In cosmology, the maximum power and maximum force arise regularly.
The maximum force $F_{\rm max}$ acting over the Hubble radius -- whose area is typically $1/\Lambda$ -- gives ${\Lambda c^4}/{8 \pi G}$.
So there is a relation between maximum force and dark energy.

The luminosity of the complete universe appears to be exactly \emph{equal} to the power limit $c^5/4G$, within measurement accuracy. 
This equality directly explains the temperature-time relation of the standard big bang scenario \cite{lastsivaram,ttbb}.
In addition, maximum power, when integrated over time, determines the critical density of the universe as an upper limit.

The electric field is defined as force per charge. 
In nature, electric charge is quantized in integer multiples of the down quark charge $-e/3$. 
As a result, a maximum force and a minimum charge imply maximum values for electric and magnetic fields given by $ E_{\rm max} = 3c^4/4Ge = 5.7\cdot 10^{62}\,\rm V/m$ and $ B_{\rm max} = 3c^3/4Ge = 1.9\cdot 10^{54}\,\rm T$.
%
However, these electromagnetic field limits cannot be tested in experiments: observable field values are limited by the much lower Schwinger limit that describes the onset of electron-positron pair production. %
The Schwinger limit is also the reason that maximum luminosity $c^5/4G$ cannot be reached by electromagnetic sources, 
     but only by sources of gravitational waves.

It should be mentioned that even {charged} black holes cannot produce forces larger than the maximum force \cite{haye}.
The black hole charge reduces the horizon radius, but the force limit among black holes or the force on test particles, whether charged or not, remains valid: the force is given by curvature, and curvature is due to all interactions combined.  %

Maximum force and power hold independently of quantum theory. 
Therefore, the two limits can be combined with quantum theory to produce additional insights and tests. 
For example, the quantum of action leads to a lower limit on time given by twice the Planck time: 
$t_{\rm min} \geq\sqrt{{4G\hbar}/{c^5}} \approx 1.1 \cdot 10^{-43}\,\rm s$,
and similarly, to a lower limit on length, given by 
$l_{\rm min} \geq\sqrt{{4G\hbar}/{c^3}} \approx 3.2 \cdot 10^{-35}\,\rm m$, and a corresponding lower limit on area and volume.
Including the quantum of action also leads to upper limits on acceleration, jerk, and density. 
No contradictions with experiments or with expectations arise \cite{MYNEWPRD}.

In 1929, Szilard \cite{szilard} argued that the smallest entropy in nature is given by the Boltzmann constant $k$.
Combining the Boltzmann constant, the force and speed limits, and the quantum of action allows deducing an upper temperature limit $T_{\rm max}=\sqrt{\hbar c^5/4G k^2}\approx 7.1 \cdot 10^{31}\,\rm K$, i.e., half the Planck temperature. %

The force limit also explains black hole entropy, discovered by Bekenstein and Hawking, again in 1973, as well as the entropy bound. 
Black hole entropy is a horizon entropy. Therefore, it is the upper limit for the entropy of a physical system enclosed by a given surface area. 
The horizon area is measured in multiples of the smallest area. 
For this reason, the factor $1/4$ from maximum force also appears in the expression $S/k= Ac^3/4G\hbar$ for black hole entropy \cite{MYNEWPRD}.

When a material reaches its elastic limit, thus when it deforms plastically before ripping, its quantum properties become important, and \textit{defects} arise. 
Similarly, just before the vacuum reaches its elastic limit at the maximum force, its quantum properties become important, and \emph{particles} arise, in the form of thermal radiation. Stretching imagination, black hole radiation can thus be seen as due to the plastic deformation of space near the horizon.

\smallskip
\section{Outlook}

\noindent In summary, general relativity can be deduced from the principle of maximum force or from the principle of maximum power, together with maximum speed. 
The search for experimental counterexamples is not successful, neither in black hole mergers, nor in astronomy, nor in cosmology. 
Also theoretical tests in the various parts of physics confirm the limits.
The limits provide several testable predictions; in particular, they suggest that general relativity is valid without any modification.
Above all, the force and power limits offer new options for teaching and exploring the fascinating beauty of general relativity.

\newpage
\smallskip
\acknowledgments

\noindent The authors thank John Donoghue, Gary Gibbons, Naresh Dadhich, Valerio Faraoni, Long-Yue Li, Michael Good, Matt Visser, Shahar Hod, Ofek Birnholtz, Barak Kol, Shahar Hadar, Pavel Krtou\v{s},  Andrei Zelnikov, Grigory Volovik, Eric Poisson and Steven Carlip for fruitful discussions.

\bibliography{main} 

\end{document}